# All-optical switching of magnetic domains in Co/Gd heterostructures with Dzyaloshinskii−Moriya Interaction

Anni Cao,[a, b] Youri L.W. van Hees,[a] Reinoud Lavrijsen,[a] Weisheng Zhao [*, b] and Bert Koopmans[*, a]

[*] Corresponding authors

[a] Department of Applied Physics, Institute for Photonic Integration, Eindhoven University of Technology, PO Box 513, 5600 MB Eindhoven, The Netherlands

[b] Fert Beijing Institute, BDBC, School of Microelectronics, Beihang University, Beijing 100191, China

Ⓢ *Supporting Information*

**ABSTRACT** Given the development of hybrid spintronic-photonic devices and chiral magnetic structures, a combined interest in all-optical switching (AOS) of magnetization and current-induced domain wall motion in synthetic ferrimagnetic structures with strong Dzyaloshinskii−Moriya Interaction (DMI) is emerging. In this study, we report a study on single-pulse all-optical toggle switching and asymmetric bubble expansion in specially engineered Co/Gd-based multilayer structures. In the absence of any external magnetic fields, we look into the AOS properties and the potential role of the DMI on the AOS process as well as the stability of optically written micro-magnetic domains. Particularly, interesting dynamics are observed in moon-shaped structures written by two successive laser pulses. The stability of domains resulting from an interplay of the dipolar interaction and domain-wall energy are compared to simple analytical models and micromagnetic simulations.

After the observation of helicity-dependent switching in ferrimagnetic GdFeCo alloys [1,2] in 2007, AOS attracted a growing amount of interest as an ultrafast and energy-saving writing process for spintronic devices. This helicity-dependent switching description has been found appropriate for various magnetic materials [3–6], however for most materials with a disadvantage that hundreds of pulses are required for the switch [6]. At the same time, based on another mechanism, purely thermal toggle switching was demonstrated in rare-earth - transition-metal (RE-TM) alloys, governed by the large difference in demagnetization rates and anti-ferromagnetic exchange [7–10]. Later, it was predicted and proved that synthetic-ferrimagnetic systems can also be thermally toggle-switched by a single laser pulse [11–14]. More specifically this was experimentally demonstrated in Pt/Co/Gd systems. Meanwhile, in the field of spintronics, an antisymmetric exchange interaction, the Dzyaloshinskii-Moriya Interaction (DMI), has been intensively investigated. It appears in inversion asymmetric structures and participates in the competition with exchange interaction and dipolar interaction to influence chiral spin textures. Considering the structural-asymmetry in synthetic-ferrimagnetic multilayers, the role of built-in DMI in the switching process, as well as the stability of the toggle-pulse-switched domains are worth to be investigated. Recently, a combination of racetrack memory [15] and AOS has been experimentally demonstrated in an 'on-the-fly' demonstration of optically writing information into a magnetic racetrack [16]. In the latter work, current-induced domain wall motion (driven by the spin Hall effect) was ascribed to a strong spin-orbit torque (SOT) in combination with DMI. For synthetic ferrimagnetic systems near the angular momentum compensation point, this scenario is known to enable high domain wall velocities [16]. Looking into the role of DMI on the AOS process is not only significant for improving the storage density and stability of such optically written racetrack devices, but also for AOS-related magnetic random-access memory (MRAM).

In this paper, we employed synthetic-ferrimagnetic Pt/Co/Gd stacks for AOS. The strong anti-ferromagnetic coupling at the Co/Gd interface [16] and the large contrast in demagnetization times between Co and Gd [17,18] ensure the optical-switching of our samples, in a scenario that recently was explained in more detail [19,20]. To manipulate the strength of DMI through structural (a)symmetry, structures with a single ferrimagnetic interface (Co/Gd) and double ferrimagnetic interface (Co/Gd/Co) are proposed. We also prepared another group of samples with different capping materials to investigate the effect of the intermixing between the capping layer and the under-layers. Asymmetric domain wall motion is observed indicating the existence of DMI. After the comparison of AOS properties, we focus on the potential role of the DMI on the field-free AOS process and the stability of optically written magnetic domains. Two successive laser pulses were produced to write elongated moon-shaped magnetic domains in our synthetic-ferrimagnetic thin-films. A simple model including the competition between dipolar interaction and domain-wall energy is proposed, which describes the domain shrinkage at the sharp ends and expansion at their waist. Both the micromagnetic simulation and experimental observations exhibit obvious shrinkage and negligible expansion. Moreover, DMI is found to stabilize the AOS-written small-size domain stripes as expected.

To distinguish the strength of DMI through structural asymmetry, we grow samples (see Figure 1 (a)) with single and double Co layers by magnetron sputtering and marked them as **S/Pt**: Ta(4 nm)/Pt(4 nm)/Co(1 nm)/Gd(3 nm)/Pt(4 nm) and **D/Pt**: Ta(4 nm)/Pt(4 nm)/Co(1 nm)/Gd(3 nm)/Co(1 nm)/Pt(4 nm), where 'S' and 'D' refers to single and double. Since time-dependent intermixing (particularly at the Gd/Pt interface) over time scales of days to weeks were observed in a similar structure, we also prepared another pair of structures with different capping layers, **S/Ta** and **D/Ta**, as reference. All of the structures are deposited on 2 types of substrates, Si:B and silicon with a 100-nm thermal silicon oxide layer (Si/SiO$_2$). The base pressure of our ultrahigh vacuum deposition system is around $2 \times 10^{-9}$ mbar. The (111) texture of the bottom Pt was ensured by a Ta seed layer [21], which promotes perpendicular magnetic anisotropy (PMA) induced by the lower Pt/Co interfaces. The PMA is confirmed by polar MOKE, as seen in Figure 1 (b).

The Curie temperature of bulk Co is 1403 K while the Curie temperature of bulk Gd (289K) lies below room temperature (298K). Considering that the Curie temperature for nanometer-thin films is even lower, the Gd layer will be paramagnetic at room temperature. However, due to the strong anti-ferromagnetic exchange interaction at the Co/Gd interface, a layer of roughly 1-2 atomic layers will be magnetized oppositely to Co at room temperature. This process is possibly enhanced by thermodynamically driven interdiffusion [22] between Co and Gd, which is expected to be of particular relevance for the top Gd/Co interface in the double-layer structures, leading to a composition at the interface close to a Gd$_{40}$Co$_{60}$ alloy [23,24]. The compensation temperature ($T_{comp.}$) is defined as the temperature for which the total magnetic moment of the whole structure vanishes. As a rule of thumb, the higher $T_{comp.}$, the more Gd is inversely magnetized at the Co/Gd interface at room temperature.

We measured the magnetic moment of our samples using the vibrating sample magnetometer (VSM) mode of a superconducting quantum interference device (VSM-SQUID) to evaluate the magnetic moment of our stacks as a function of temperature, and, more specifically, $T_{\text{comp.}}$, from which we can extract the amount of magnetized Gd at the synthetic-ferrimagnetic Co/Gd interfaces [13]. For each sample, we applied + 6 T external field to saturate the film at the start, then turn off the field, and measure the perpendicular component of the magnetic moment at different temperatures. The magnetic moments per unit surface area versus temperature are shown in Figure 1 (c). While the temperature decreases, the magnetization of the Gd layer increases, and below $T_{\text{comp.}}$, the Gd, instead of the Co layer, starts to dominate the total moment. Upon further decrease of the temperature, the total magnetic moment will rapidly increase (in the direction of the Gd moment), whereby the (in-plane) shape anisotropy will overcome the (perpendicular) interfacial anisotropy, and starts to be dominant. As a result, the easy axis will rotate from out of plane to in-plane, as seen in the measurements by a sharp collapse of the magnetic moment for lower temperature (indicated by pointers in Figure 1 (c)). More details about measuring $T_{comp.}$ with a bias field and the corresponding temperature-dependent magnetic moments are given in Supporting Information part I.

For a proper interpretation, it is of relevance to note that for structures with two Co layers and a double Co/Gd interface, one would ideally expect both a doubling of the Co moment, as well as the Gd moment. Thus, based on the assumption that the top (Gd/Co) and bottom (Co/Gd) interface are their exact mirror image, one would expect that the $T_{comp.}$ of double Co/Gd interface structures will be comparable with $T_{comp.}$ of the single (Co/Gd) interface group. However, as seen in Figure 1 (c), **S/Ta** shows a lower $T_{comp.}$ (177 K) and a higher room-temperature moment compared with **D/Ta**, which means that the top and bottom interfaces are not identical. More specifically, the higher $T_{comp.}$ of **D/Ta** would be consistent with a larger induced Gd moment at the top interface. This finding agrees with work by T. C. Hufnagel *et al.*, who reported that compared with the case when Co is deposited on Gd, less rapid intermixing is observed when Gd is deposited on Co [22]. The same trend is observed for the samples capped with Pt, albeit they display a higher $T_{comp.}$ compared to their Ta counterparts, and the difference between **S** and **D** structure is larger. For structures with a single Co layer, the sample capped with Pt (**S/Pt**) exhibits less intermixed Gd compared with **S/Ta**, which might be caused by the intermixing between Gd and the Pt capping layer. Furthermore, note that we did not observe $T_{comp.}$ below room-temperature for **D/Pt** in Figure 1 (c). However, the inverse Kerr signal, shown in Figure 1 (b), indicates an inverse total moment of the whole structure which means the Gd moment is larger than the Co moment in **D/Pt**. Therefore, we deduce that **D/Pt** has a $T_{comp.}$ above room-temperature indeed.

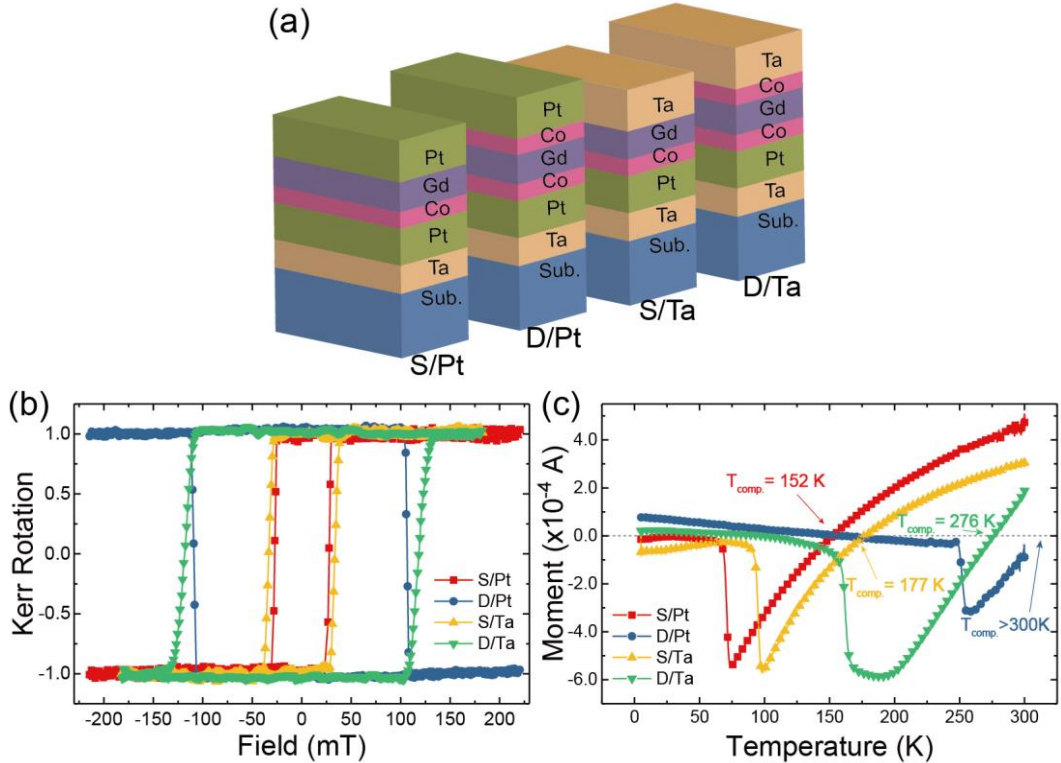

Figure 1. (a) Samples' structures and (b) normalized hysteresis loops with perpendicular magnetic field for each structure grow on Si/SiO$_2$. (c) Magnetic moment per unit area as a function of the temperature, with $T_{comp.}$ corresponding to the zero-moment marked besides each curve.

We quantified the strength of DMI in our samples employing a magneto-optical Kerr microscope to observe asymmetric DW movement in the creep regime with an in-plane field $H_x$ and a perpendicular field $H_z$. The dependence of DW velocities on the in-plane field is found to display a minimum occurring at a non-zero value of $H_x$, which we use as an indication of the strength of DMI and mark it as $H_{DMI}$ [25]. A typical result of asymmetrical DW motion in the presence of an in-plane field $H_x$ is shown in Figure 2 (a). The applied $\mu_0 H_z$ is around tens of milli-tesla, and the in-plane field $\mu_0 H_x$ is in the range of ± 350 mT. Due to strong pinning and $T_{comp.}$ being close to room-temperature, we failed to observe DW motion in sample **D/Pt**. The other three samples displayed a very pronounced asymmetry in the DW velocity which indicates a sizable DMI. We employed a procedure using an empirical function (see Supporting Information part II) by which we obtained the values of the effective DMI field $H_{DMI}$, as shown in Figure 2 (b) – (d). In a ferromagnetic system, the DMI energy can be extracted by $|D| = \mu_0 M_S |H_{DMI}| \sqrt{A/K_{eff}}$ [26], where $M_S$ is the saturation magnetization, $A$ the exchange stiffness and the $K_{eff}$ effective anisotropy. As we mentioned above, in a Co/Gd system, there is a very thin layer of Gd antiferromagnetically coupled with Co at the Co/Gd interface. It has been reported that the thickness of the antiferromagnetic Gd layer is about 0.5 nm. Although extracting a quantitative value of the DMI parameter $D$ is far from trivial for this system, we performed a simplified analysis to derive a rough estimate. By assuming the thickness of ferromagnetic layer $t$ = 1.5 nm for a single Co structure (**S/Pt** and **S/Ta**) and $t$ = 3.0 nm for a double Co structure (**D/Pt** and **D/Ta**), $M_S$ and $K_{eff}$ (averaged over Co and Gd per unit volume) can be calculated based on the VSM data. Taking $A$ = 16 pJ/m from literature [27], we estimate $|D| \approx$ 0.09±0.01, 0.24±0.01 and 0.37±0.01 mJ/m$^2$ for **D/Ta**, **S/Ta** and **S/Pt** respectively. We do notice that the characterization of DMI by the minimum DW velocity in the creep mode has been intensively debated. Nevertheless, employing the same measurement and a model including the most important contributions to the domain-wall energy, Hartmann *et al.* have reported a $H_{DMI}$ ~ 220 mT in Pt/Co/Gd structure [25], which is close to our result of **S/Ta** as shown in Figure 2 (c). This correspondence may indicate that for our Co/Gd structures creep data provides a reasonable and trustworthy value of the DMI field. However, independent of such a quantitative comparison, the strong asymmetries we measured in the DW velocities are a clear indication of a significant DMI for all samples, and a reduced DMI of the double structure as compared to the single one shown in Figure 2 (c) and (d). This is consistent with our expectation that the double-Co-layer structure should be more symmetric, i.e. have a smaller $H_{DMI}$, because of the inversion symmetry of the Co/Gd and Gd/Co interface. The observation that **D/Ta** still has a finite $H_{DMI}$, hints again at the non-equivalence of the bottom and top interface, as could be explained by growth-induced differences. We stress that our estimate of $|D|$ is just a rough estimate for several reasons, including but not limited to the uncertainty of $t$ and $A$. Hereon, we leave the exact value of $|D|$ in our Co/Gd system as an open question for further research.

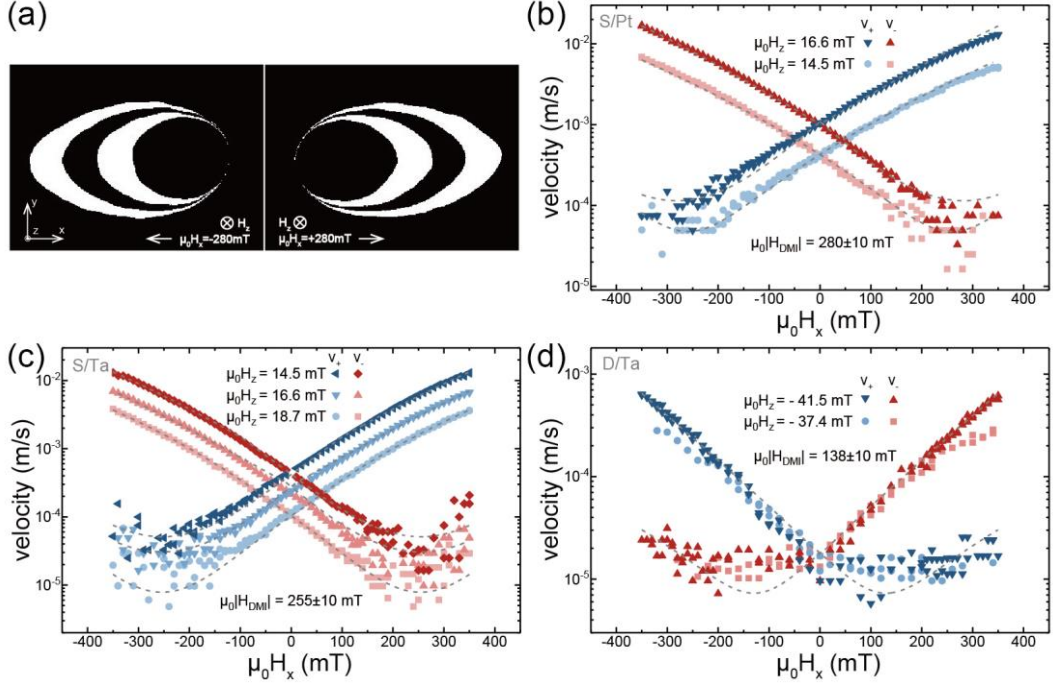

Figure 2. (a) DW expansion of **S/Ta** driven by out-of-plane magnetic field $\mu_0|H_z| = 16.6$ mT with in-plane field $\mu_0|H_x| = 280$ mT. DW velocities of structure **S/Pt**, **S/Ta** and **D/Ta** are shown in (b), (c) and (d) respectively, with $\mu_0|H_x|$ varies from -350 mT to 350 mT and corresponding $\mu_0 H_z$ as well as the fitted $\mu_0|H_{DMI}|$ (corresponding to the gray dashed lines) are indicated. $v_+$ (blue symbols) present the domain's velocity along the +x axis, $v_-$ (red symbols) present the domain's velocity along the -x axis.

We carried out all-optical switching experiments on these synthetic-ferrimagnetic multilayers. All samples were firstly saturated with an external field and then excited by single laser pulses without applying a magnetic field. Each structure was exposed to linearly polarized laser pulses with 700 nm central wavelength, ~100 fs duration, and a pulse energy up to 1 µJ, focused on typically a spot size of ~50 µm diameter. The all-optically written domains were imaged after mounting the samples in a Kerr microscope. In this section, we will discuss experiments aimed at exploring the AOS efficiency of the various samples. In the following sections will report on experimental investigations aimed at understanding the stability of small, optically written micro-magnetic domains, after first having provided some physical insight by introducing some simple models.

As outlined above, we start by exploring the AOS efficiency. A threshold fluence $F_0$ is defined as the threshold pulse energy $P_0$ divided by the laser spot area $A_0$ (defined at $1/e$ intensity). Considering the thermal switching mechanism, demagnetization of the magnetic layer is a precondition for magnetization switching. We varied the pulse energy by a neutral density filter wheel and wrote circular domains by single laser pulses, shown in Figure 3 (a). It can be seen that above the threshold energy $P_0$, the domain area exhibits a positive correlation with pulse energy. The areas of domains written on the four samples are plotted in Figure 3 (b) as a function of pulse energy. Supposing the energy of the laser pulse shows a Gaussian spatial profile, in the case of an elliptical laser spot, the relation between the domain area and the pulse energy has been derived to be [28]

$$A = 2\pi r\sigma^2 \ln\left(\frac{P}{P_0}\right) = A_0 \ln\left(\frac{P}{A_0 F_0}\right), \qquad (1)$$

where $\sigma$ is the length of the short axis for the elliptical Gaussian spot and $r$ the ratio between the long and short axis. $r$ and $\sigma$ were determined from the Kerr images, after which $A_0$ and $F_0$ can be fitted using Equation (1). The fitted $F_0$ of the four structures grown on Si/SiO$_2$ are smaller than $F_0$ for structures with Si:B substrate, due to more efficient optical absorption. More details are discussed in Supporting Information part III.

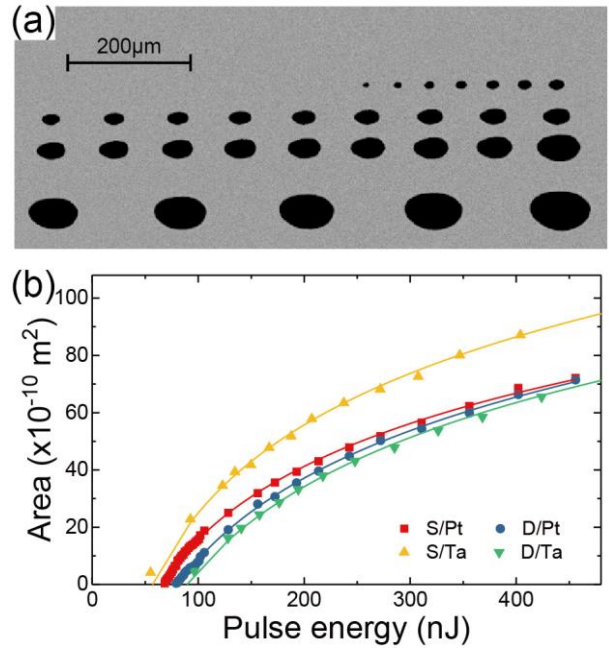

Figure 3. Pulse-energy-dependent AOS measurement on Co/Gd stacks. (a) Kerr image of the switching experiments performed on **S/Pt**. An increasing domain area is observed upon increasing the pulse energy up to 450 nJ. (b) Domain size as a function of pulse energy for different structures. The minimum pulse energy to nucleate a domain (intercept of x-axis) corresponds to the threshold energy $P_0$.

To understand the stability of small domains after AOS, we will distinguish two types of domains, as can be simply produced by one or two laser pulses. (i) Circular bubbles are written by AOS and using a single pulse, such as shown in Figure 3 (a). Note that these domains are expected to form skyrmionic bubbles, i.e., Néel-type domain walls with net chirality at zero fields when the DMI is strong enough. (ii) Moon shape patterns are produced by two successive, slightly displaced laser pulses, which converge to extremely thin, curved stripes if the displacement of the laser pulses is much smaller than the diameter of the laser spot, see Figure 4 (a). The latter one (type (ii)) will be discussed in the following part. We propose that the small domains display shape changes due to the laser's heating effect or the following creep processes. To describe the creep-like deformation of the narrow-stripe-domain, we successively address two parts: the sharp tip and the middle region.

For an ideal writing process, governed by an entirely local switching event with a well-defined threshold fluence, two moon-shaped domains are formed, touching in two singularities at opposite sides of the laser spot, where two domain walls cross. Such a structure is intrinsically unstable, as we verified by micro-magnetic simulations below. At a time-scale of nanoseconds, the two touching domains will detach as illustrated in Figure 4 (a), ending up in two tapered domains with rounded ends. To further reduce their micromagnetic energy, these tapered ends will shrink like Figure 4 (b) illustrates. This shortening process is dominated by the reduction of domain wall energy to lower the total energy, where the domain wall energy density is defined as $\sigma = 4\sqrt{AK_{eff}} - \pi D$ (per unit area) [29,30]. In this way, not only $A$ and $K_{\text{eff}}$ but also the strength of DMI is related to the instability (shrinkage) of the domain at its tip end; the larger the DMI energy the larger its stability, so the slower the shrinkage.

To make an order of magnitude estimate of the effective field governing the shrinkage process, we approximate the tip end by a sharp triangle with a semi-circular cap, where $\theta$ is the opening angle, $w$ is the diameter, and $t$ is the ferromagnetic layer thickness (see Figure 4 (b)). In the limit of $\theta \to 0$, this structure can be considered as a stripe with fixed-width terminated by a semi-circular cap. For this structure, the domain wall energy will reduce by $dE = 2\sigma t dx$ when the stripe contracts by $dx$. Neglecting the change in the dipolar field associated with this contraction, and comparing with the difference of Zeeman energy in an external field ($\Delta E = -\boldsymbol{m} \cdot \boldsymbol{B}$), the effective field corresponding to the reduction of DW energy can be written as $B_{\text{eff}}^{\text{DW}} = \frac{2\sigma t dx}{w t M_s dx} = \frac{2\sigma}{w M_s}$. Note that upon contraction, and assuming a small but finite value of $\theta$, the value of $w$ will increase, leading to a gradually slowing down of the shrinkage process.

However, this contraction will not continue indefinitely. When the width of the stripe $w$ grows larger, dipolar energies, which strive to increase the stripe width, will increase. In order to obtain a simple estimate, we approximate the domain by a semi-infinite stripe with a straight cap. To avoid divergencies in the integral for dipolar energies, the domain wall width $\Delta$ should be assumed to be finite. According to the Biot-Savart's Law and the difference of energy, we derived an approximate equation for the effective field associated with the dipolar interaction: $B_{\text{eff}}^{\text{di}} = \frac{\mu_0 M_s t}{\pi w} \ln\left[\frac{(w-\Delta)^2}{\Delta^2}\right]$, where $\Delta$ is the width of DW (see Supporting Information part IV). To be noticed, $B_{\text{eff}}^{\text{di}}$ has the opposite sign as $B_{\text{eff}}^{\text{DW}}$. As a consequence, the total effective field of the shortening process can be expressed as

$$B_{\text{eff}}^{\text{sh}} = \frac{2(4\sqrt{AK_{\text{eff}}} - \pi D)}{w M_s} - \frac{\mu_0 M_s t}{\pi w} \ln\left[\frac{(w-\Delta)^2}{\Delta^2}\right]. \tag{2}$$

Note that, as a consequence, $B_{\text{eff}}^{\text{sh}}$ will fade-out slowly with increasing $w$.

Finally, we note that near the center of the moon-shaped domain, it can be modelled as a stripe of width $w$ with infinite length, corresponding to assuming $\theta \to 0$. In this limit, the broadening of the stripe at its center does not cost any domain wall energy, but dipolar interaction would result in a lower total energy with an increasing $w$. As a consequence, in this limit, there will always be a tendency to broaden. In practice, this process stops because the effective field becomes too small to lead to measurable creep, or the domain transforms into a shape with a less pronounced aspect ratio, in which the net driving force to reduce the DW energy may result in a net shrinkage everywhere, ultimately leading to a collapse of the complete domain. Seen in Figure 4 (c), an expansion process driven by dipole interaction will be observed in the middle of the narrow domain. Regarding the infinite length of the stripe domain (compared with its width), we can obtain the effective field

$$B_{\text{eff}}^{\text{ex}} = \frac{2\mu_0 M_s t}{\pi w}. \tag{3}$$

Thus, it is found that the larger the domain width $w$, the smaller the driving field associated with the expansion. With a specific $w$ and experimental $M_s t$, $B_{\text{eff}}^{\text{ex}}$ of each structure can be quantitively obtained. For our structures, with a width of at least 1 micro-meter, and using typical parameters, this leads to effective fields of 0.4 mT or smaller. In order to make an estimate of the relevance of such an effective field for our samples, we measured the domain wall velocity as a function of perpendicular field (at zero in-plane field). We fitted the results for each sample using the relation between the creep velocity and the driving field $\ln(v) \sim (\mu_0 H_z)^{-\frac{1}{4}}$, and extrapolated the trends according to these fits (more details can be found in Supporting Information part V). Thus, our estimate of $B_{\text{eff}}^{\text{ex}}$ extrapolates to a maximum velocity less than 1 μm in several days for all samples, which means that in our case the expansion process is expected to be negligible. We note that this weak tendency of widening a domain at its center is related to the small magnetization of our films, as a consequence of the compensating Co and Gd moments, leading to small dipolar effects.

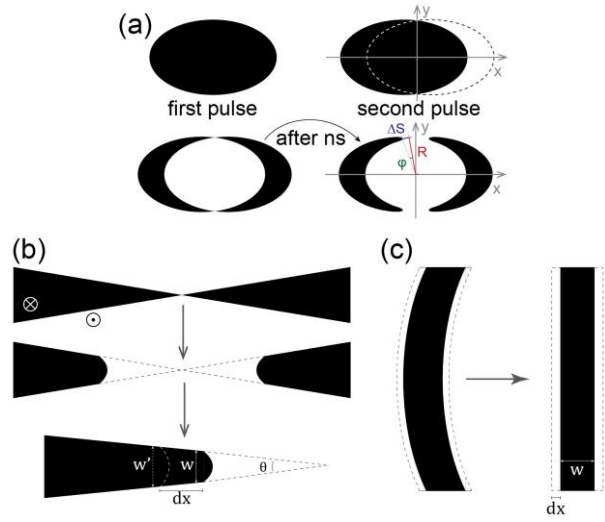

Figure 4. (a) A schematic of moon-shape domain pairs writing by 2 toggle laser pulses. (b) Shortening process of domains on the sharp tips. (c) Widening process of domain strip.

We simulated the shape-change process of the moon like domain pair by the micromagnetic package Mumax3 without applying any magnetic field and at $T = 0$. A series of results can be seen in Figure 5, where domain pairs with different 'step sizes' (i.e. the displacement of the laser spot between firing the two successive pulses) share the same time scale marked at the right. The first line of Figure 5 (a) - (c) corresponds to the relaxed state of each domain pair. We found that the spontaneous shortening process happens at a ns time scale after the creation of the domain. The shrinking velocity damps rapidly, in agreement with Equation (2), while the smaller the step size the larger the shrinking velocity. Besides, there isn't any clear expansion process during the short-term time scale for all of the step sizes in our simulation. Finally, we note the asymmetry in shrinkage at the top and bottom end of the domain, particularly for the smallest step size (Figure 5 (a)). This artefact is due to a combination of our finite mesh size and the point like anomaly in the initial state with crossing domain walls, but resolving it goes beyond the scope of this paper.

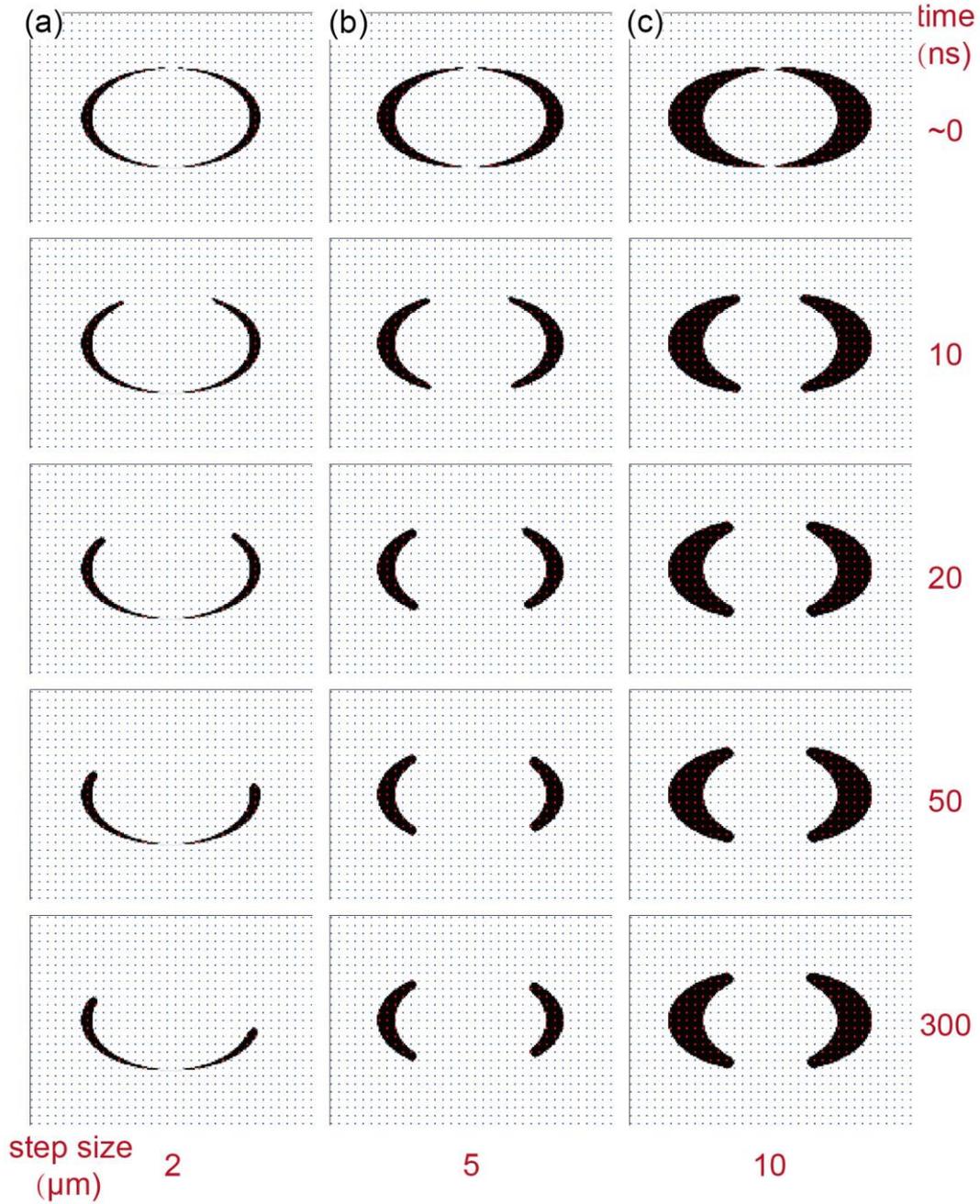

Figure 5. Time-dependent shape-change of domains simulated by Mumax3, with a step size of (a) 2 μm, (b) 5 μm, and (c) 10 μm.

After understanding the main driving forces for the creep-like modifications of the AOS-written domains, the experimental results are discussed in this section. We generated moon-shaped domain-pairs by two consecutive laser pulses with a shift (step size) of 10 μm, 5 μm, 2 μm, and 1 μm. All samples were saturated by ~ 56 mT external field before the optical switching. Kerr images for switching results of different samples are shown in Figure 6, sharing the scale bar shown in Figure 6 (d). All of the images were obtained within 5 min after AOS. Due to having its $T_{comp.}$ above room-temperature, **D/Pt** shows an inverse Kerr signal and correspondingly white domain areas, as compared to the black domains for the other three structures. It can be seen that the sharpest parts in the top and bottom ends of each domain-pair have eclipsed in a similar fashion as introduced in Figure 4 (a), (b) and Figure 5. This demonstrates that an initial shortening process happened in the time interval between creating the domains and measuring the Kerr images. We also notice that the thinner domains (corresponding to largest $\theta$ and smallest $w$) produced with small step size suffer from stronger quenching processes, while more stable domains are produced with larger step sizes. We decreased the step sizes until the domains are unstable even at a time scale of minutes. For the **S/Pt**, **D/Pt**, **S/Ta**, and **D/Ta** structures, this led to a minimum step size of (roughly) 2 μm, 1 μm, 2 μm, and 5 μm respectively. According to Equation (2), the effective driving field of the shortening process $B_{\text{eff}}^{\text{sh}}$ has a contribution that is proportional to the DW energy, which is negatively correlated to the DMI energy $D$. Comparing the domain images at 2 μm step size in Figure 6 (c) and (d), the sample with larger structural asymmetry (**S/Ta**) still shows some small domains, whereas all domains were completely quenched for the **D/Ta** sample. This behavior hints at a larger stability for domains in **S/Ta** structures, which is consistent with Equation (2). Since **D/Pt** exhibit $T_{\text{comp.}}$ higher than room temperature, other complicated mechanisms might be involved in the domain's stability, we refrain from a discussion of the group capped by Pt (**S/Pt** and **D/Pt**) here.

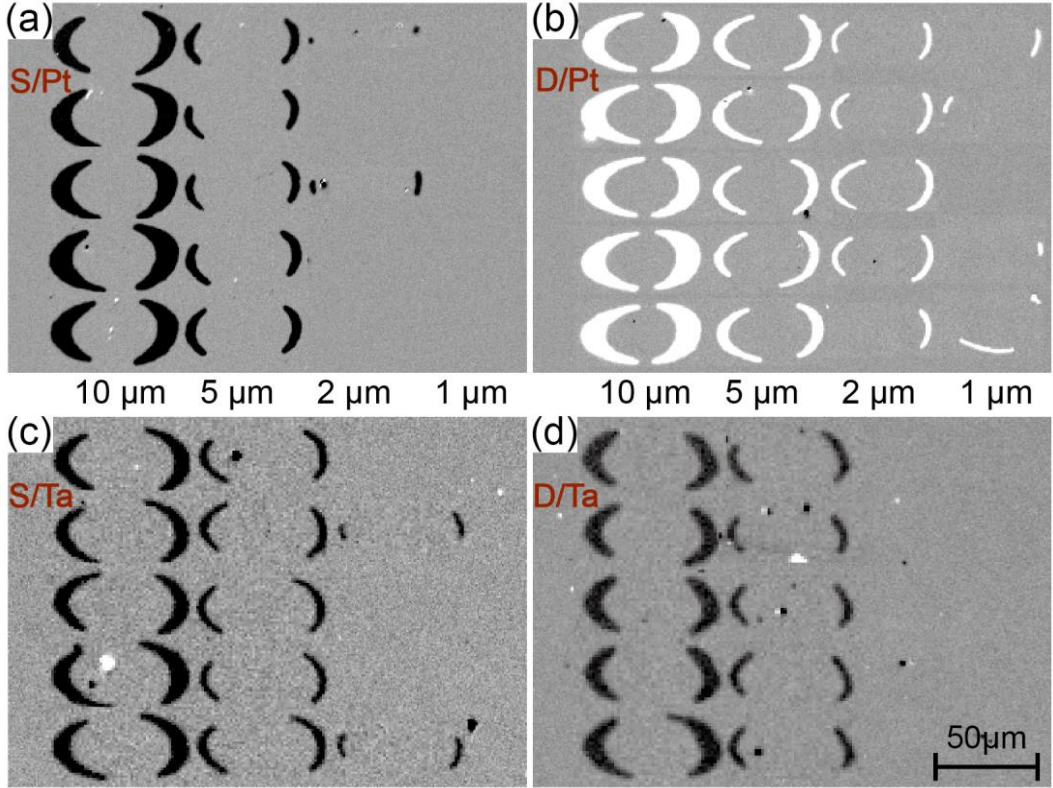

Figure 6. The moon-shape domain pairs writing by 2 toggle laser pulses on the sample (a) **S/Pt**, (b) **D/Pt**, (c) **S/Ta**, and (d) **D/Ta**. Different step sizes result in different domains' width, while each column from left to right corresponding to step size 10 μm, 5 μm, 2 μm, and 1 μm. Scale bar is shown in (d).

To describe the shape-change of domains more quantitatively and with specific emphasis on domain dynamics at a longer time scale, we analyzed the variance of the waist width and the opening angle, corresponding to the expansion and shrinkage respectively. The definitions of the width and the opening angle ($\lambda_l$ and $\lambda_r$) are given in Figure 7 (a). In passing we note that the observed difference between $\lambda_l$ and $\lambda_r$ can be explained by an asymmetry of the laser spot. For that reason, we averaged $\lambda_l$ (left domain) and $\lambda_r$ (right domain) of each domain pair with the same step size. Furthermore, we averaged width of domains within ± 5° range around the x-axis. The averaged $\lambda$, as well as the averaged width, as a function of time (after AOS) for each structure are shown in Figure 7 (b) – (d).

We first focus on a possible widening effect. The dashed black lines in the top panels of Figure 7 (b) - (d) represent the averaged width of the ideal shape (without expansion), which is determined by the corresponding step sizes. Taking the error bars into account, the measured (averaged) width does not significantly deviate from the ideal values, i.e., there is not any observable expansion process in both the short-term and the long-term time scales after AOS.

Next, we address the shortening process, by inspecting the lower panels of Figure 7 (b) – (d). Right after AOS, the initial value of $\lambda$ (the average of $\lambda_l$ and $\lambda_r$) should be 180°. The measured values are clearly smaller, corresponding to the initial shrinkage that we already qualitatively discussed before. Beyond this initial shrinkage that happens at a short time scale after AOS, the values of λ display a clear further reduction at a time scale of up to an hour. We fitted the trends of the averaged λ from 3 to 60 minutes by $\lambda = A - Bt$. The fitted parameter $B$ is proportional to the shortening velocity at this longer time scale. Converted to domain wall velocities, values up to approximately $10^{-9}$ - $10^{-8}$ m/s are found. Detailed values are given in Supporting Information part V. The bottom panels in Figure 7 (b) - (d) show that domains generated at smaller step size shorten faster than domains at larger step size, which is consistent with the simulation results shown in Figure 5.

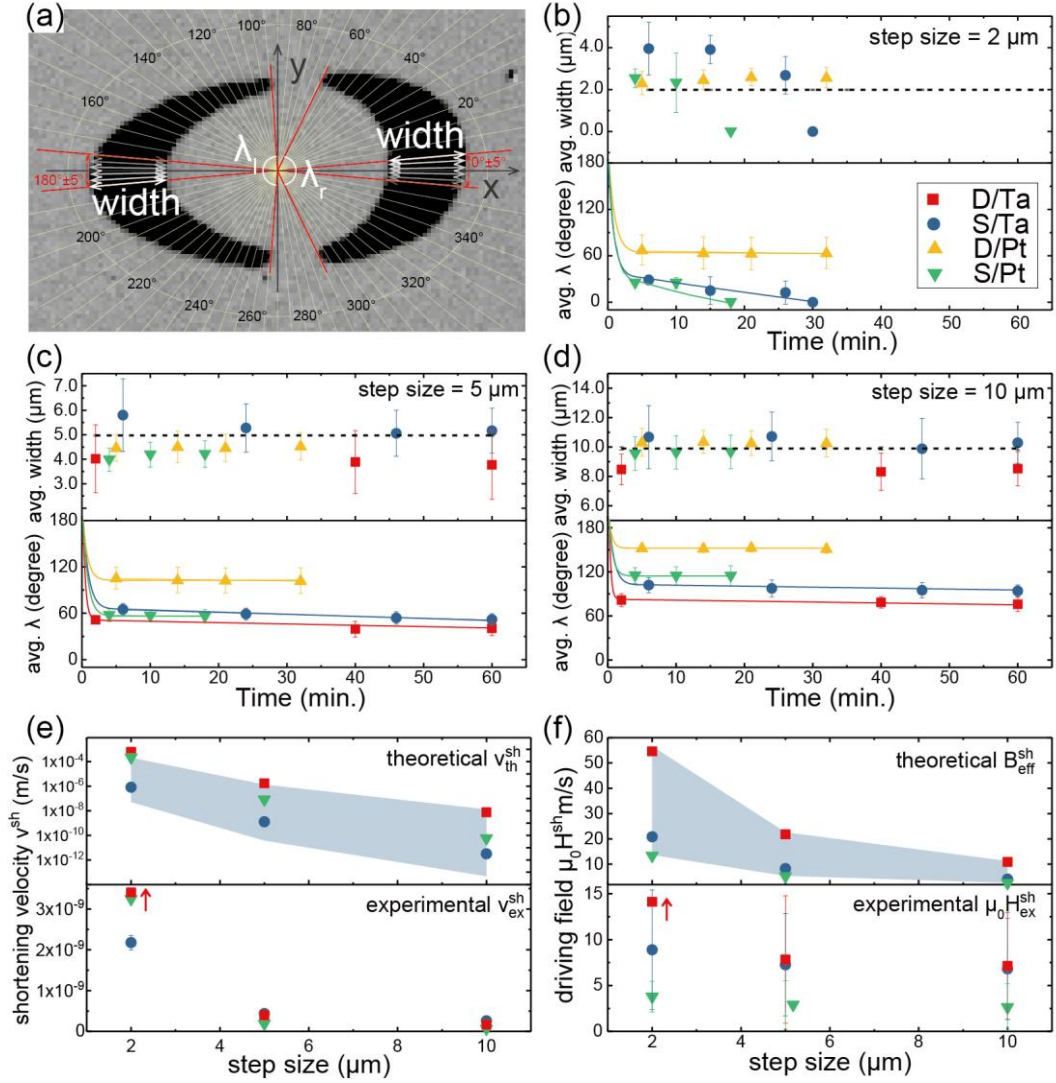

Figure 7. (a) Kerr image of a domain-pair on **D/Ta** stacks, including the definition of width and λ in the later figures. (b) - (d) the time evolution of average width (top) and average λ (bottom) for domains written at 2 μm, 5 μm, and 10 μm step size during tens of minutes after AOS. Comparison of (e) shortening velocity and (f) driving field between experimental and theoretical data. Due to the uncertainty of the ferromagnetic layer thickness, we take **S/Ta** as an example, calculate confidence intervals for $v_{th}^{sh}$ and the corresponding $B_{eff}^{sh}$ shown in (e) and (f) as light blue zones. Note that at 2 μm step size all domains on **D/Ta** were quenched before the first Kerr image was taken, indicative for a large shortening velocity and driving field, as schematically indicated by the red arrow. (b) - (f) share a common legend shown in (b).

Based on the above experimental results (Figure 7 (b) - (d)) and physical model (Equation (2)), we quantitatively compared the theoretical shortening velocity $v_{th}^{sh}$ with the experimental shortening velocity $v_{ex}^{sh}$, as well as the corresponding driving fields ($B_{eff}^{sh}$ and $\mu_0 H_{ex}^{sh}$). Results are shown in Figure 7 (e) and (f), respectively. **D/Pt** is omitted in this part due to the absence of reliable DMI measurements. First, we discuss how we derived the 'experimental' values. We use the fitted parameter $B$ to calculate $v_{ex}^{sh}$. In the limit of $\varphi \to 0$, $\theta \approx \tan\varphi$, and thus $v_{ex}^{sh} = \frac{\Delta S}{\Delta t} = \frac{R}{2}\frac{\Delta \lambda}{\Delta t}$, where $R$ equals to the radius of a domain written by a single laser pulse (~ 25 μm in our case), $\Delta S$ stands for the shortening length within a period of time $\Delta t$, and $\frac{\Delta \lambda}{\Delta t}$ corresponds to the fitted value of $B$. The definition of $\Delta S$, $R$, and $\varphi$ can be found in Figure 4 (a). Extrapolating the measured field-induced DW velocities using the creep law, we also estimated the experimental driving field $\mu_0 H_{\text{ex}}^{\text{sh}}$. The error bar of the experimental data comes from the standard error of the fitted $B$. Next, we address how we derived the 'theoretical' values. The driving field for shortening can be obtained from Equation (2). $B_{\text{eff}}^{\text{sh}}$ can be calculated by adopting a value of exchange stiffness $A = 16$ pJ/m from the literature [27], using the saturation magnetization $M_S$ and effective anisotropy $K_{\text{eff}}$ from our experimental data (VSM data, assuming $t = 1.5$ nm or 3.0 nm), DW width $\Delta = \sqrt{A/K_{\text{eff}}}$, DMI energy $D$ (as we introduced in the DMI section) and domain width $w$. To be noticed, we approximated the domain width $w$ by the step size values to simplify the calculations, but verified that the error range on the domain width $w$ does not significantly affect the order of $B_{\text{eff}}^{\text{sh}}$. Similarly, the theoretical shortening velocity $v_{\text{th}}^{\text{sh}}$ can be calculated by extrapolating our DW velocity data according to the creep law. Taking **S/Ta** as an example, we change the ferromagnetic layer thickness $t$ from 1 nm to 4 nm (2 nm to 5 nm for D/X structure), recalculate $M_S$ (per volume), $K_{\text{eff}}$ (per volume), $D$, then obtained confidence intervals for $v_{\text{th}}^{\text{sh}}$ and $B_{\text{eff}}^{\text{sh}}$, shown as the light blue zones in Figure 7 (e) and (f). Detailed values of each parameter as well as the relation between the creep velocity and the driving field are provided in Supporting Information part V.

Comparing the experimental and theoretical estimates in Figure 7 (e) and (f), a reasonable agreement can be concluded on. (i) Within order of magnitude the driving fields compare well; both the experiment and theoretical predictions yield fields of the order of 10 mT. Although due to the exponential behavior the theoretical velocities vary over many orders of magnitude, also there some overlap can be observed. (ii) We found the same trend in shrinkage as a function of domain width. Both the theoretical and experimental results show that the shrinkage is more pronounced in narrower domains. (iii) As to the role of DMI, at 2 μm step size, a clear stabilizing effect of DMI is found in the Ta-capped set of samples. A lower DMI led to a quenching of all domains. For the data at larger step sizes, the experimental velocities and driving fields show a less systematic effect. The effect of DMI is certainly smaller than the theoretical values predict. As to the other discrepancies between theory and experiments, in particular the observation that the absolute values of $v_{\text{th}}^{\text{sh}}$ and $B_{\text{eff}}^{\text{sh}}$ are appreciably larger than $v_{\text{ex}}^{\text{sh}}$ and $\mu_0 H_{\text{ex}}^{\text{sh}}$, they can be attributed to several reasons. Firstly, as described by our physical model on the shrinkage, under the limitation of $\theta \to 0$, the length change of the semicircular cap $\Delta w$ (from $w$ to $w'$ shown in Figure 4 (b)) can be neglected. In the case of a specific $\theta$, the contribution of $\Delta w$ would result in a smaller $\frac{2(4\sqrt{AK_{\text{eff}}} - \pi D)}{wM_S}$ (positive term of Equation (2)) and a smaller $B_{\text{eff}}^{\text{sh}}$. Secondly, with a specific $\theta$, the dipole interaction would be larger, thus result in a larger $\frac{\mu_0 M_S t}{\pi w} \ln \left[\frac{(w-\Delta)^2}{\Delta^2}\right]$ (negative term of Equation (2)) and also a smaller $B_{\text{eff}}^{\text{sh}}$. Thirdly, the estimations of the DMI energy and the ferromagnetic layer thickness are quite rough. Finally, taking the light blue areas (confidence interval) in Figure 7 (e) and (f), as well as the error bars into account, we conclude that our model is in reasonable qualitative agreement with the experiments.

In conclusion, we carried out a series of experiments on Pt/Co/Gd multilayers with single and double Co/Gd interfaces, which in the ideal case should correspond to asymmetric and symmetric structures, respectively. Firstly, the compensation temperature was measured to confirm the synthetic-ferrimagnetic behavior of these stacks. Then the DMI strength was estimated by measuring asymmetric field-induced DW motion in the presence of an in-plane field. Next, we proposed a physical model for two types of shape-change in narrow domain stripes, i.e., a shortening and a widening process. Micromagnetic simulations and experimental observation on the optically written domains were also performed, which qualitatively and quantitively confirmed the shrinkage process and excluded the expansion process of domains. Through the comparison between the theoretical and experimental results on the shrinkage of domains, we demonstrated the applicability of our model, which describes the role of DMI on the shortening process of optically switched domains. More specifically, we conclude that DMI as inherently built into the Pt/Co/Gd stacks, is helpful for stabilizing AOS-written small-size domain stripes. We also notice that it is not trivial to scale down to sub-micron dimensions as would be required for large data densities, which can be another significant follow-up research to do.

**AUTHOR INFORMATION**

**Correspongding Author**

*E-mail: b.koopmans@tue.nl (B. K.) and weisheng.zhao@buaa.edu.cn (W.-S. Z.).

**Acknowledgment.** The authors thank the National Key R&D Program of China 2018YFB0407602, the National Natural Science Foundation of China (61571023), the International Collaboration Project B16001, the National Key Technology Program of China 2017ZX01032101, the Program of Introducing Talents of Discipline to Universities in China (No. B16001), the VR innovation platform from Qingdao Science and Technology Commission and the China Scholarship Council.